\documentclass[sigchi]{acmart} 

\usepackage{booktabs} 
\usepackage{changepage} 
\usepackage[inline]{enumitem}
\usepackage{balance}

\newcommand{\deem}[1]{\textcolor{gray}{\small#1}}

\copyrightyear{2019} 
\acmYear{2019} 
\setcopyright{acmlicensed}
\acmConference[CHI 2019]{CHI Conference on Human Factors in Computing Systems Proceedings}{May 4--9, 2019}{Glasgow, Scotland UK}
\acmBooktitle{CHI Conference on Human Factors in Computing Systems Proceedings (CHI 2019), May 4--9, 2019, Glasgow, Scotland UK}
\acmPrice{15.00}
\acmDOI{10.1145/3290605.3300429}
\acmISBN{978-1-4503-5970-2/19/05}

\graphicspath{{./figures/}}

\begin{document}
\title[Understanding Perceptions of Problematic Facebook Use]{Understanding Perceptions of \\ Problematic Facebook Use} 
\subtitle{When People Experience Negative Life Impact and a Lack of Control}

\author{Justin Cheng}
\affiliation{%
  \institution{Facebook}
}
\email{jcheng@fb.com}

\author{Moira Burke}
\affiliation{%
  \institution{Facebook}
}
\email{mburke@fb.com}

\author{Elena Goetz Davis}
\affiliation{%
  \institution{Facebook}
}
\email{elenadavis@fb.com}

\begin{abstract}

While many people use social network sites to connect with friends and family, some feel that their use is problematic, seriously affecting their sleep, work, or life.
Pairing a survey of 20,000 Facebook users measuring perceptions of problematic use with behavioral and demographic data, we examined Facebook activities associated with problematic use as well as the kinds of people most likely to experience it.
People who feel their use is problematic are more likely to be younger, male, and going through a major life event such as a breakup.
They spend more time on the platform, particularly at night, and spend proportionally more time looking at profiles and less time browsing their News Feeds.
They also message their friends more frequently.
While they are more likely to respond to notifications, they are also more likely to deactivate their accounts, perhaps in an effort to better manage their time.
Further, they are more likely to have seen content about social media or phone addiction.
Notably, people reporting problematic use rate the site as more valuable to them, highlighting the complex relationship between technology use and well-being.
A better understanding of problematic Facebook use can inform the design of context-appropriate and supportive tools to help people become more in control.

\end{abstract}

%
%



\maketitle

\section{Introduction}

Social network sites help people maintain social relationships \cite{ellison2007benefits,burke2016relationship}, drive civic engagement and collective action \cite{gil2012social,obar2012advocacy}, and support entrepreneurship \cite{harris2009social}. But while many people derive benefit from online social networks, some feel that their use of such services is problematic. Studies of problematic use of the internet (e.g., \cite{cash2012internet,young1998internet}) and social networks (e.g., \cite{andreassen2012development,kuss2017social,ryan2014uses}) note symptoms including preoccupation, loss of control, and negative impact on one's relationships, work performance, and life \cite{griffiths2005components}.

The present study focuses on perceived problematic Facebook use to understand its prevalence and its relation to different activities on the site, in order to inform design improvements that may reduce problematic use.
We define ``problematic Facebook use'' as reporting a significant negative impact on sleep, relationships, or work or school performance and feeling a lack of control over site use, consistent with broad definitions from the academic literature \cite{peng2010social,ryan2014uses}.
We do not use the term ``addiction'' because there is no agreed-upon criteria for diagnosis \cite{tao2010proposed,griffiths2012facebook,billieux2015we}, and because diagnoses of clinical-level concerns would require more formal assessment (i.e., by a mental health professional) \cite{ko2009proposed}.
Instead, we focus on self-reported problematic use to understand differences across a broad population of users.

We pair a survey of 20,000 Facebook users in the U.S. measuring perceived problematic Facebook use with server logs of aggregated behavioral data for the previous four weeks, such as the amount of time respondents spent on the site and counts of interactions with close friends. In contrast to prior work that has relied on small-sample, survey-only analyses of problematic social network use \cite{ryan2014uses}, mostly among adolescents and young adults (e.g., \cite{andreassen2012development,durkee2012prevalence}), we use a larger, more diverse sample to study how perceptions of problematic use relate to actual on-site activity. By drawing data from both surveys and server logs, we reduce common-method bias, in which problematic outcomes and self-reported time online appear more associated than they are in reality \cite{podsakoff2012sources}.

Under this broad definition of problematic Facebook use -- negative life impact and difficulty with control -- we
estimate (as an upper bound) that 3.1\% of Facebook users in the US experience problematic use. They are more likely to be younger, male, and going through a major life event such as a breakup. After controlling for demographics, we find that people experiencing problematic use spend more time on the platform, particularly at night, and respond to a greater fraction of notifications. Contrary to stereotypes of people scrolling through endless content, people who experience problematic use spend proportionally less time in their News Feeds and more time browsing profiles, and message others more frequently. People reporting problematic use are also 2.6 times as likely to deactivate their accounts, perhaps as a way to control the time they spend on the site. They are also more likely to have viewed posts and comments about social media or phone addiction. And despite feeling that their use of the site has a negative impact in their lives, they rate Facebook as more valuable to them than do people in the non-problematic use group.

\section{Background}

First, we review literature on problematic internet use and problematic Facebook use, and identify open questions about how individual differences and behaviors relate to the latter.

\subsection{Problematic Internet and Facebook Use}
Problematic internet use has been described as a set of symptoms including excessive amounts of time spent on the internet, a preoccupation with online activities or inability to control one's use, and adverse impact on one's social interactions and work or school performance \cite{cash2012internet}. Though academic researchers have described problematic internet use empirically, no formal clinical definition exists in either the Diagnostic and Statistical Manual of Mental Disorders \cite{american2013diagnostic} or the International Classification of Diseases \cite{world2018international}. There is also disagreement on whether such behaviors comprise a defined disorder \cite{griffiths2012facebook} and whether research is pathologizing common behaviors \cite{billieux2015we}. Moreover, previous surveys that attempt to measure problematic internet use (e.g., \cite{young1996psychology}) have adopted inconsistent assessment criteria, leading to widely differing prevalence estimates \cite{kuss2014internet}. These estimates have ranged from 1.5\% to 8.2\% in the US and Europe \cite{weinstein2010internet}, to 0.3\% to 38\% internationally \cite{chakraborty2010internet}. Nonetheless, there has been substantial academic and clinical interest in researching problematic internet use and related issues such as problematic Facebook use, problematic online gaming, and nomophobia (a fear of being out of mobile phone contact) \cite{bragazzi2014proposal,kuss2017social}.

While there is debate on how problematic Facebook use should be measured \cite{andreassen2012development,lee2012investigation,griffiths2012facebook} (or if it should be classified as an addiction \cite{billieux2015we}), a majority of survey instruments (e.g., \cite{andreassen2012development}) include questions about lack of control, or a failure to abstain from the activity, and negative life impact, such as relationship conflict or reduced work or school performance \cite{griffiths2005components,kuss2011online}. Other proposed symptoms from the behavioral addiction literature include salience, or how much one thinks about or engages in site use; tolerance, or needing increasing amounts of activity over time to achieve a desired effect; and mood modification and withdrawal, defined as a reliance on site use to reduce unpleasant feelings \cite{griffiths2005components,ryan2014uses}. Still, researchers have argued against using these symptoms as diagnostic criteria because of an absence of clinical studies \cite{kardefelt2017can}. Measures of symptoms such as tolerance that are adapted from diagnostic criteria for substance abuse may also not be appropriate when applied to technology use \cite{billieux2015we}.
Consistent with survey instruments used in prior literature, this work focuses on problematic use as self-reporting both significant negative life impact and difficulty controlling Facebook use.

Prior literature suggests that symptoms of problematic internet use may be due to co-occuring problems \cite{pies2009should} -- individuals with problematic internet use tend to have other psychiatric disorders \cite{kratzer2008internet}.
Past research has associated problematic internet or Facebook use with depression \cite{kim2006internet}, lower happiness \cite{brailovskaia2018physical}, worse academic performance \cite{kirschner2010facebook}, greater loneliness \cite{ryan2011uses}, and reduced relationship and life satisfaction \cite{elphinston2011time,blachnio2016association}, though null results have also been reported \cite{biolcati2018facebook}.
Problematic internet behaviors may also arise from other individual differences.
Previous work suggests that a preference for online social interaction may contribute to using the internet in problematic ways, especially when a person feels lonely or depressed \cite{caplan2003preference}, has low self-esteem \cite{andreassen2017relationship,kanat2018contingent}, or is neurotic or narcissistic \cite{andreassen2012development,mehdizadeh2010self}. A fear of missing out (``FOMO''), more formally defined as ``a pervasive apprehension that others might be having rewarding experiences from which one is absent'' \cite{przybylski2013motivational}, might also contribute to problematic smartphone, internet, and social media use \cite{oberst2017negative,kuss2017social,buglass2017motivators}. Problematic internet use has also been associated with structural differences in the brain \cite{he2017brain}.

\subsection{Demographic Differences Related to Problematic Use}

\subsubsection{Gender}
Evidence is mixed regarding whether men or women are more likely to experience problematic internet use.
Previous work found that women tend to use Facebook more than men \cite{foregger2008uses}, and some studies indicate that women are more likely to experience problematic use \cite{andreassen2012development,de2014compulsive,banyai2017problematic} or report communication disturbance and phone obsession \cite{blachnio2018aware}.
However, other studies showed a higher prevalence of problematic internet use among men \cite{ccam2012new,durkee2012prevalence,yen2009multi}.
Other work found no significant relation between gender and problematic internet use \cite{blachnio2016association,rumpf2014occurence,tang2016personality}.

\subsubsection{Age}
Past research suggests that younger people may be more likely to experience problematic use because regions of the brain responsible for self-regulation are still developing in adolescence \cite{steinberg2008social} and because they are more susceptible to negative peer influence \cite{steinberg2007age}.
Other work also found that internet addiction negatively correlates with age \cite{fernandez2015validation}.
Correspondingly, a majority of previous studies of problematic use focus on these younger subpopulations (e.g., adolescents \cite{ko2005gender,banyai2017problematic} or college students \cite{fernandez2015validation,koc2013facebook}).
In the present work, we survey a wide range of Facebook users in the U.S. to better understand the relationship of both gender and age across a larger sample of people.

Because the existing literature is mixed on the relationship between gender and problematic use, and because little research has been done on problematic use across a wide range of ages, we pose the following research question:

\vspace{2mm}
\begin{adjustwidth}{0mm}{}
\noindent\textit{RQ1: How does problematic use differ by gender and age?}
\end{adjustwidth}

\subsection{Behaviors Associated With Problematic Use}

Previous literature has also examined how specific behaviors relate to perceptions of problematic Facebook use. We discuss three main themes across behaviors:
\begin{enumerate*}
    \item excessive time spent,
    \item connections and tie strength, and
    \item loss of control.
\end{enumerate*}
We also briefly examine the role of social narratives in shaping individual perceptions of problematic use.

\subsubsection{Excessive time spent}
Previous work has correlated time spent with both problematic internet  \cite{durkee2012prevalence,fernandez2015validation} and problematic Facebook use \cite{koc2013facebook,hong2014analysis}. Greater time spent has also been associated with social anxiety \cite{shaw2015correlates}. However, spending extended periods of time on the internet or on Facebook does not necessarily suggest problematic use \cite{caplan2005social}. Whereas generalized problematic internet use may involve displacing social connections, past research suggests that greater Facebook use may support people's relationships, depending on how they use it \cite{burke2016relationship,deters2013does,verduyn2015passive}. Moreover, research has also found that people spend substantial time on online social networks to maintain their offline social networks \cite{kuss2011online}, and people who use Facebook several times a day have more close ties than people who do not \cite{hampton2011social}. Some work found a quadratic relationship between well-being and time spent online, with moderate use associated with improved well-being \cite{przybylski2017large,twenge2018decreases}.

Previous research has also linked problematic Facebook use to late-night use.
It has been associated with both later bedtimes and rising times \cite{andreassen2012development} and with insomnia \cite{koc2013facebook,uysal2013mediating}.

Overall, the relationship between time spent on Facebook and problematic use is unclear. Thus, we ask:

\vspace{2mm}
\begin{adjustwidth}{0mm}{}
\noindent\textit{RQ2: How does time spent relate to problematic use?}
\end{adjustwidth}

\subsubsection{Connections and tie strength}
Past work has associated problematic Facebook use with poorer well-being \cite{ryan2014uses,brailovskaia2018physical}, so indicators of well-being may correlate negatively with problematic use. In particular, a large body of research suggests that interacting with close friends can lead to improvements in well-being, more so than interacting with acquaintances \cite{wellman1990different,bessiere2008effects,burke2016relationship,valkenburg2007online}. If a person spends much of their time on Facebook interacting with acquaintances rather than close friends, this could influence their evaluation of the quality of the time she spends on the site, and their overall determination of whether their use is problematic. While little work on problematic internet use has focused on its relation to tie strength in online interactions, prior work noted higher levels of upward social comparison among people with more acquaintances in their Facebook friend graph \cite{chou2012they}. Offline, people are more likely to underestimate others' difficult moments and overestimate their successes \cite{jordan2011misery}, and the effect online could be stronger among acquaintances, who may be less likely to know of each other's negative emotions. Thus, one might expect that if a person's Facebook network were denser and consisted of a greater ratio of strong to weak ties, they might experience improvements in well-being that buffer any negative impact from Facebook use.

Problematic use may also be associated with differences in friending and messaging behavior, but research is mixed. On one hand, individuals with low self-esteem may engage in friending more actively to compensate for a perceived deficiency \cite{lee2012wants}. On the other hand, teens who used Facebook to make friends reported reduced loneliness \cite{teppers2014loneliness}. Prior work has associated instant messaging use with positive outcomes such as increased intimacy and improved relationship quality through increased self-disclosure \cite{hu2004friendships,valkenburg2009effects} and with negative outcomes such as problematic use \cite{van2008online}. As such:

\vspace{2mm}
\begin{adjustwidth}{0mm}{}
\noindent\textit{RQ3: How do interactions with close friends on Facebook relate to problematic use?}
\end{adjustwidth}

\subsubsection{Loss of control}
Survey measures of problematic use commonly include lack of control \cite{cash2012internet}. We focus on two categories of Facebook activities that affect control: notifications and deactivation. Notifications may prompt people to use Facebook at times when they wouldn't have otherwise, thus reducing feelings of control by interrupting other tasks or in-person social interactions. In prior work, interruptions slow task completion \cite{czerwinski2000instant}, inhibit performance on complex tasks \cite{speier2003effects}, and make it difficult to return to a previously interrupted task \cite{o1995timespace}. Previous research also found that notifications can cause inattention and hyperactivity, which in turn decreases productivity and subjective well-being \cite{kushlev2016silence}.

\vspace{2mm}
\begin{adjustwidth}{0mm}{}
\noindent\textit{RQ4: How do notifications differ between people who experience problematic use and those who don't?}
\end{adjustwidth}
\vspace{2mm}

Deactivation, or temporarily disabling one's account, is another method of control. Past work suggests that people may deactivate to focus during periods of high stress (e.g., before an exam), when they feel they spend too much time on Facebook, or to prevent others from interacting with their content while they are not online \cite{baumer2013limiting,boyd2010risk}.
\vspace{2mm}
\begin{adjustwidth}{0mm}{}
\noindent\textit{RQ5: How do deactivation patterns relate to problematic use?}
\end{adjustwidth}

\subsubsection{Social narratives}
Previous research has shown that what people read or hear about can influence their beliefs \cite{deutsch1955study,mccombs2002agenda}.
Reading an op-ed can result in substantial, long-term shifts in a person's policy opinions \cite{coppock2018long}.
Further, previous qualitative work found that social narratives about smartphone addiction and its negative consequences can lead to people perceiving their own smartphone use negatively \cite{lanette2018much}.
\vspace{2mm}
\begin{adjustwidth}{0mm}{}
\noindent\textit{RQ6: How does reading about social media or smartphone addiction relate to perceptions of problematic use?}
\end{adjustwidth}

\section{Method}

To measure the prevalence of problematic Facebook use and the behaviors associated with it, we surveyed Facebook users in May 2018 and combined survey responses with server logs of the participants' activity on Facebook in the four weeks prior to them taking the survey. To protect participants' privacy, all data were de-identified, aggregated, and analyzed on Facebook's servers; no identifiable data were viewed by researchers. An internal board reviewed the research prior to the start of the study.

\subsubsection{Participants}
Participants (N=20,505; 62\% female; mean age 44.5) were recruited via an ad on Facebook targeted at a random sample of people in the U.S. Compared to active Facebook users, respondents were on average 3.6 years older, 15\% more likely to be female, had 20\% more friends, and had owned their Facebook accounts for 1 year longer (all comparisons \textit{p} < 0.001). To account for these differences, as well as general differences in site use due to demographics, we control for age, gender, friend count, and account tenure in all behavioral analyses below.

\subsubsection{Problematic use survey}
The survey contained questions about control and negative life impact adapted from the Internet Addiction Test \cite{young1998internet}, the Generalized Problematic Internet Use Scale 2 \cite{lee2012investigation}, and the Bergen Facebook Addiction Scale \cite{andreassen2012development}, see Table \ref{tab:survey}. The survey also asked ``How valuable do you find the time you spend on Facebook?'', ``How meaningful are your interactions with people on Facebook?'', and whether the respondent experienced any major life events in the past two months: `moved to a new city', `relationship breakup or divorce', `lost job', `new job', `pregnancy or new family member', `death of close friend or family', or `personal injury or illness' \cite{holmes1967social}. Participants opted in to taking the survey; the survey stated ``Some of these questions may be sensitive and all are optional; you may prefer not to answer and that's okay.''

\begin{table*}
  \renewcommand*{\arraystretch}{1.05}
  \begin{tabular}{p{\textwidth}}
    \toprule
    \textbf{Survey items measuring problematic use} \\
    \midrule
    \multicolumn{1}{l}{\textbf{Negative life impact}} \\
    How often do you get less sleep than you want because you're using Facebook? \\[-0.2em]
    \deem{Never / Rarely / Sometimes / Very often / Always} \\[0.2em]
    Overall, how much does your use of Facebook hurt your relationships with others? \\[-0.2em]
    \deem{Very slightly or not at all / A little / Moderately / Very much} \\[0.2em]
    To what extent does Facebook help or hurt your work or school performance? \\[-0.2em]
    \deem{Helps greatly / Helps somewhat / Neither helps nor hurts / Hurts somewhat / Hurts greatly} \\[0.2em]
    Overall, do you feel like Facebook has had a positive or negative impact in your life? \\[-0.2em]
    \deem{Very negative impact / Somewhat negative impact / Neither positive nor negative impact / Somewhat positive impact / Very positive impact} \\[0.2em]
    \multicolumn{1}{l}{\textbf{Control or preoccupation}} \\
    How much control do you feel you have over the amount of time you spend on Facebook? \\[-0.2em]
    \deem{Very little or no control / A little control / Some control / A lot of control / Complete control} \\[0.2em]
    How concerned are you about missing important posts on Facebook if you don't log in frequently enough? \\[-0.2em]
    \deem{Not at all concerned /  A little concerned / Somewhat concerned / Very concerned / Extremely concerned} \\
    \bottomrule
  \end{tabular}
  \caption{The problematic Facebook use survey included questions on negative life impact and control.}
  \label{tab:survey}
\end{table*}

\subsubsection{Defining problematic use}
Based on the literature reviewed above, we defined problematic use as reporting both of the following:

\begin{enumerate}[leftmargin=*]
    \item \textbf{Negative life impact} attributed to Facebook:
    \begin{itemize}[label={$\boldsymbol{\cdot}$},align=left,labelsep=0.23cm,leftmargin=0cm]
        \item Facebook hurts their relationships ``very much,'' or
        \item They ``very often'' or ``always'' get less sleep because of Facebook, or
        \item Facebook hurts their work or school performance ``greatly,'' or
        \item Facebook has a ``very negative'' impact on their lives
    \end{itemize}
    \item and \textbf{problems with control or preoccupation}:
    \begin{itemize}[label={$\boldsymbol{\cdot}$},align=left,labelsep=0.23cm,leftmargin=0cm]
        \item ``Very little or no control'' over the time they spend on Facebook, or
        \item ``Very'' or ``Extremely'' concerned about missing posts from not logging in frequently enough
    \end{itemize}
\end{enumerate}

We require both components in our definition because voluntary choices (e.g., staying up late to use Facebook) may not be problematic if people still feel in control of those choices. We intentionally define this construct broadly, in contrast with stricter definitions proposed in the literature that require symptoms across multiple domains of functioning \cite{andreassen2012development}. Thus, our estimate is likely to be an upper bound on the prevalence of problematic Facebook use, and more likely reflects risk of problematic use.

\subsection{Measures of Potential Excessive Use}

\subsubsection{Time spent}
We include the total amount of time participants spent on the site over the past four weeks as well as the number of distinct sessions because checking frequently throughout the day may indicate habitual behavior. A session is defined as distinct if it starts at least 60 seconds after a previous session ended. Similarly, we divided each day into 24 ``hour bins'' and counted the number of distinct bins in which a person had at least one session. Because of the association between problematic use and lack of sleep, we include the fraction of sessions that occur late at night (12 - 4 AM). We also include one measure from the survey in this analysis: how valuable respondents feel their time on Facebook is.

Furthermore, feelings of problematic use may be associated with how people spend their time on Facebook, so we include the proportion of time in News Feed (where they read friends' content and provide feedback), in Messenger (where they can have private conversations), on profiles, in groups, watching videos, and on Pages (which represent small businesses, public figures, or entertainment). Time in each of these areas was divided by the total time spent.

\subsection{Measures of Connection and Tie Strength}

\subsubsection{Interactions with close friends}
We defined ``close friends'' as a respondent's top 10 Facebook friends in terms of the number of mutual friends, among people with at least 100 friends. A similar measure that defined closeness based on communication frequency and photo co-tags produced qualitatively similar results. In this study we measured the fraction of News Feed posts that respondents viewed that were produced by close friends, the fraction of messages to and from close friends, the fraction of profiles they viewed that were close friends, and their ego network density (or local clustering coefficient).

To additionally understand non-friend interactions, we measured the number of new friend requests sent in the past four weeks and the fraction of requests that were accepted, the fraction of non-friend profile views, and the fraction of comments written on non-friend posts.

\subsubsection{Messaging}
We include the number of messages participants sent in the past four weeks. All data were counts; no message text was viewed by researchers. Because the overall number of messages sent could simply be a proxy for overall time spent, we also included a normalized version, messages sent in four weeks divided by time spent on the site over four weeks.
We also examined whether people sent more messages than they received from their friends. Because a person could send many short messages or fewer longer ones, we also included average message length.

\subsubsection{Feedback received and given}
Feedback in the form of likes and comments is a common component of social media models of tie strength \cite{gilbert2009predicting,burke2014growing}. As feedback has been associated with improvements in well-being \cite{burke2016relationship}, it may also be linked to decreased problematic use. Thus, we measure the number of likes and comments received in the past four weeks normalized by the number of posts the participant wrote. We also measure the number of likes and comments participants gave, normalized by the number of posts they viewed.

\subsection{Measures Related to Control}

\subsubsection{Notifications}
We looked at the total number of push notifications received over the past four weeks, the fraction of notifications that a person responded to, and the mean time to response (in cases where there was a response).

\subsubsection{Deactivation}
We include whether the person deactivated their Facebook account in the four weeks prior to the survey. To see the survey, participants had to be currently active on the site, so these deactivations were temporary.

\subsection{Measuring The Role of Social Narratives}
To test if reading about social media or smartphone addiction is associated with feelings of problematic use, we analyzed posts and comments that participants viewed in the past four weeks, computing the fraction of posts and comments that included words relating to addiction (e.g., ``addicted'', ``compulsive'') as well as words relating to either social media or smartphones (e.g., ``Facebook'', ``phone'').
All analyses were done on de-identified data in aggregate; no post or comment text was viewed by researchers.

\subsection{Demographic Variables}
We include demographic variables including age and gender identity as covariates in our analyses, which are likely to affect both an individual's Facebook use and their perceptions of problematic use. We also include their friend count as a proxy for overall site engagement, and their account tenure in days, to control for demographic differences based on when a person joined Facebook.

\subsection{Method of Analysis}
To understand how different experiences on Facebook are associated with reports of problematic use, we divided survey respondents into two groups: those who experience problematic use based on the definition above, and those who do not. For interpretability, we report results primarily as the relative differences in the means between the two groups based on a matched sample on age, gender, friend count, and account tenure (e.g., ``people in the problematic use group spent 21.6\% more time on Facebook than people in the non-problematic use group, all else being equal''). We performed coarsened exact matching \cite{iacus2012causal}, followed by a linear regression on the matched sample to compute the average treatment effect. To account for multiple comparison, we report Holm-corrected \textit{p}-values. Comparing groups using logistic regressions controlling for age, gender, friend count, and account tenure on the entire data set (not matched samples) produces qualitatively similar results.

This method is correlational, so we cannot determine the causal relationship between survey measures of perceived problematic Facebook use and activities on the site. However, much of the existing research in this space is also correlational. By identifying associations, we can outline potential design implications and areas where additional research is needed to identify the causal direction.

\section{Results}

Here, we examine the types of people that report problematic use, and explore how activity on Facebook relates to problematic use with respect to
\begin{enumerate*}
    \item potential excessive use,
    \item connections and tie strength,
    \item a loss of control, and
    \item social narratives about addiction.
\end{enumerate*}

\subsection{Who experiences problematic use?}
Based on our definition for problematic use -- experiencing a negative life outcome attributed to Facebook as well as a lack of control -- 3.1\% of Facebook users in the U.S. experience problematic use. This estimate has been weighted by age, gender, and time spent to account for selection bias among survey participants. Because of a lack of consensus in prior literature about how to define problematic use, we include the two most common criteria -- a negative life outcome and lack of control. This is less restrictive than some models (e.g., those that require multiple negative life outcomes, mood modification, or tolerance). Therefore, our estimate of 3.1\% is an upper bound compared to other definitions with stricter criteria, but this broader definition allows us to make design recommendations more broadly.

\subsubsection{Age}
\begin{figure}
\includegraphics[width=\columnwidth]{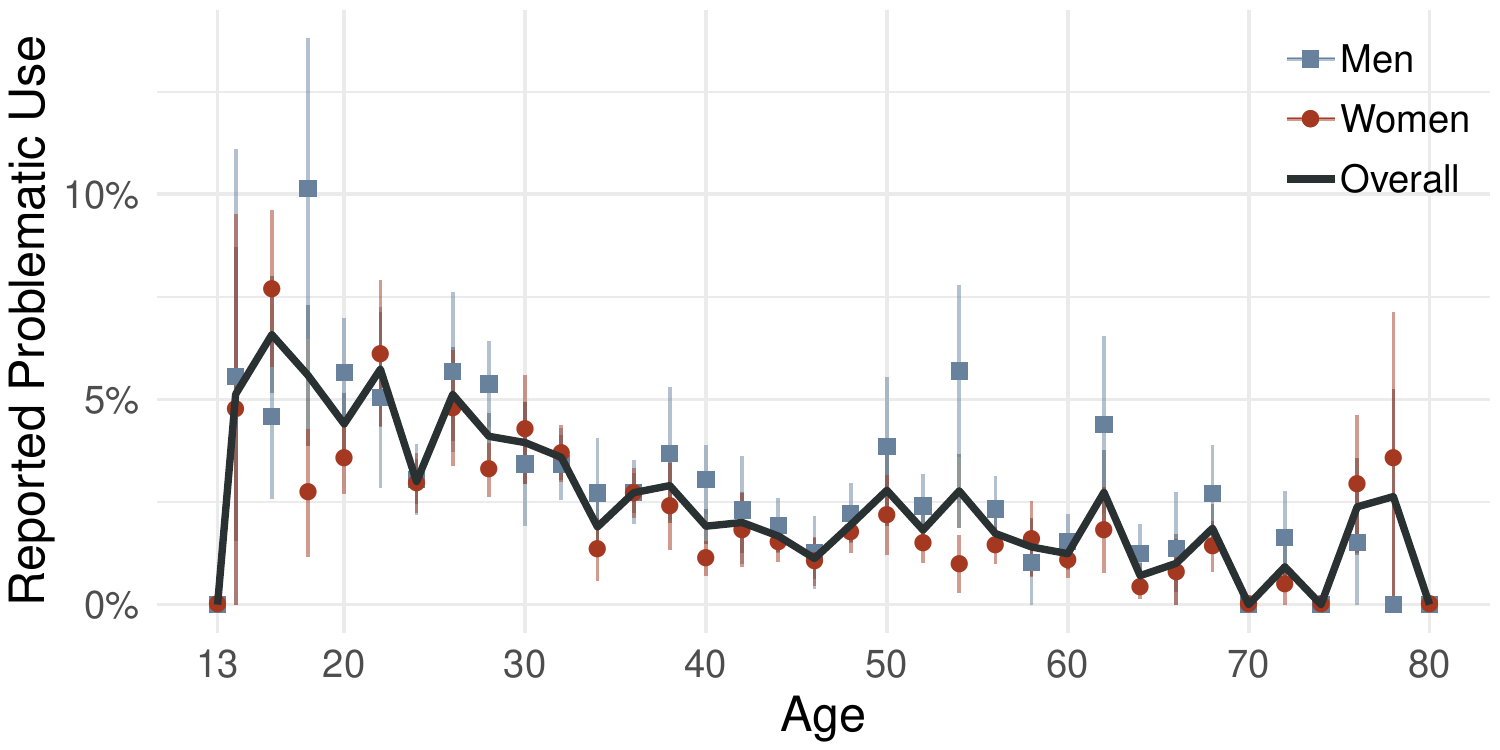}
\caption{The prevalence of reported problematic use is highest among teens and young adults. Men are also more likely than women to report experiencing problematic use.}
\label{fig:puagegender}
\end{figure}

Answers to Research Question 1, the prevalence of problematic use by age and gender, are presented in Figure \ref{fig:puagegender}. Perceptions of problematic use vary by age, with the prevalence highest among teens and young adults. People under the age of 25 were almost twice as likely as other age groups to experience problematic use (Cohen's \textit{d} = 0.13, \textit{p} < 0.001).
This is consistent with previous research showing that younger people have more difficulty with self-regulation \cite{steinberg2008social} and thus may be more prone to problematic use.

\subsubsection{Gender}
Men are 1.4x as likely as women to report experiencing problematic use (\textit{d} = 0.05, \textit{p} < 0.001, Figure \ref{fig:puagegender}). Still, the women in our sample spent 16\% more time than the men on Facebook, suggesting that the relationship between time spent and problematic use is likely mediated by other factors, including motivations for use \cite{roberts2014invisible}.

\subsubsection{Major life events}
\begin{figure}
\includegraphics[width=\columnwidth]{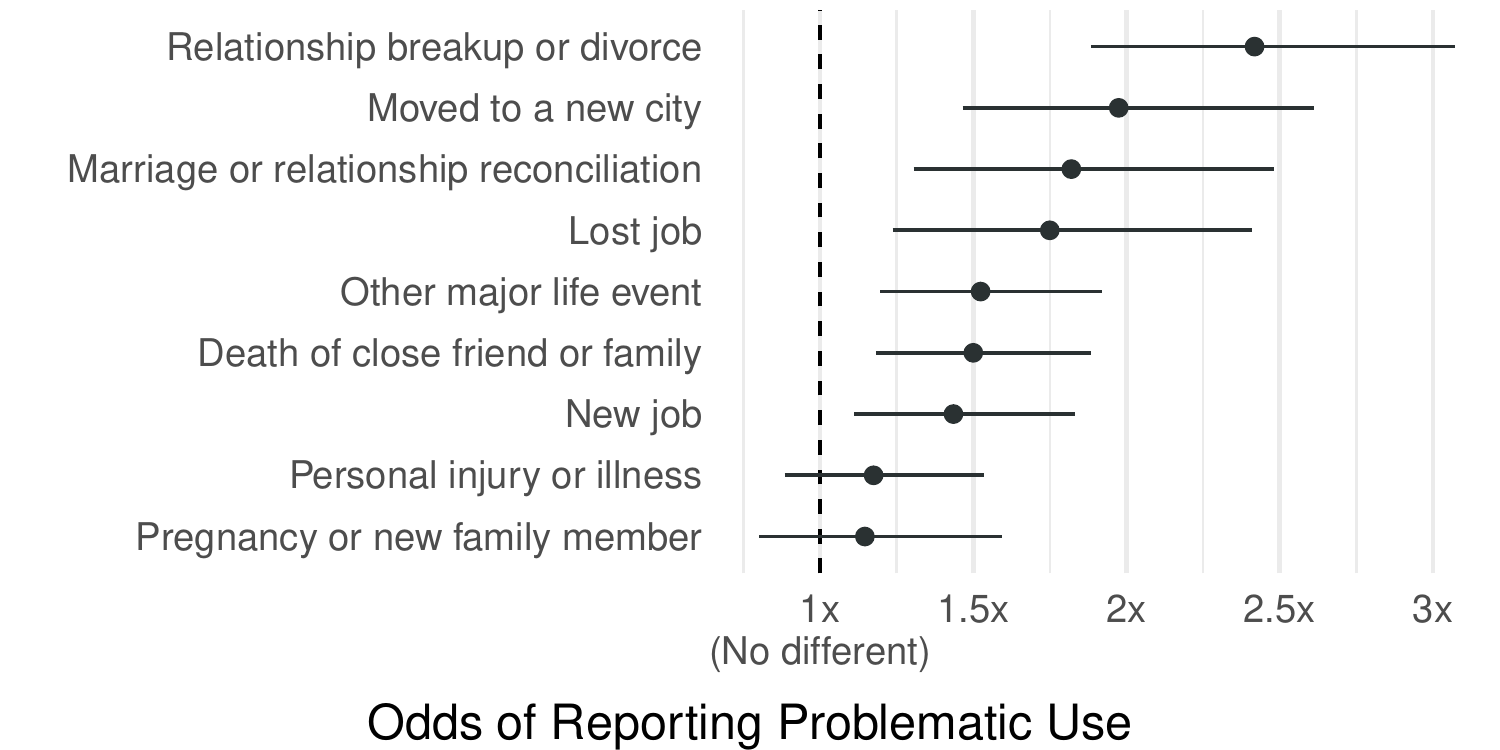}
\caption{Having a major life event (e.g., a breakup) in the past two months is associated with an increased likelihood of reporting problematic use.}
\label{fig:mle}
\end{figure}

Participants reported on the survey major life events that had happened in the past two months, and many of them were significantly associated with perceived problematic use (Figure \ref{fig:mle}). People who had recently gone through a breakup were 2.4x as likely to report that their use of Facebook was problematic. Similarly, a person who had recently moved to a new city was approximately 2x as likely to report problematic use.

\subsubsection{Friend count and account tenure}
Participants who reported experiencing problematic use had 29\% more friends (\textit{p} < 0.001), and had owned their Facebook accounts for about ten fewer months (\textit{p} < 0.001). We control for these variables in all subsequent analyses.

\subsection{Potential Excessive Use}

\subsubsection{Time spent}
\begin{figure}
\includegraphics[width=\columnwidth]{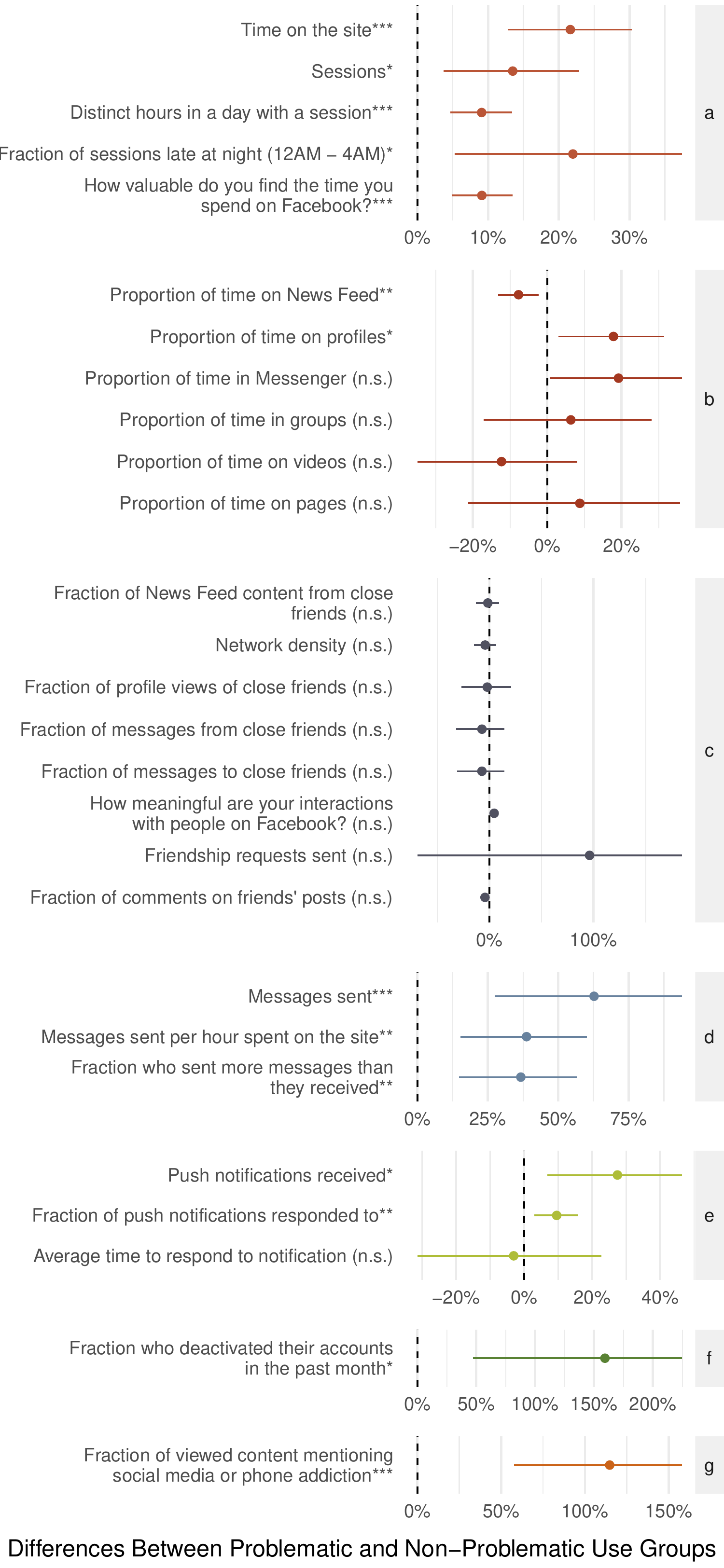}
\caption{Relative differences between people who report experiencing problematic use and those who do not, matched on age, gender, friend count, and tenure. For example, people reporting problematic use spent 21.6\% more time on Facebook than people who did not. Bars represent 99\% bootstrapped confidence intervals. All \textit{p}-values are Holm-corrected to account for multiple comparisons. \\
* \textit{p} < 0.05, ** \textit{p} < 0.01,  *** \textit{p} < 0.001,  n.s. not significant}
\label{fig:cmp_all}
\end{figure}

Consistent with prior literature, people who reported problematic use spent significantly more time on Facebook than people who did not (Figure \ref{fig:cmp_all}a, Research Question 2). They spent 21.6\% more time on the site (\textit{d} = 0.28, \textit{p} < 0.001), had 13.5\% more distinct sessions (\textit{d} = 0.16,  \textit{p} < 0.05), and had sessions in more distinct ``hour bins'' each day (\textit{d} = 0.24, \textit{p} < 0.001). They also spent a greater fraction of their sessions late at night (\textit{d} = 0.16, \textit{p} < 0.001), consistent with their increased likelihood of reporting sleep problems.

Despite reporting problems that they attributed to their Facebook use, individuals in the problematic use group found the time they spent on Facebook as 9.1\% more valuable than people in the non-problematic use group (\textit{d} = 0.24, \textit{p} < 0.001). One interpretation is cognitive dissonance: a person justifies the extra time he or she spends on Facebook by thinking that it is more valuable. However, as we later show, there were no differences between groups in how meaningful they rated their interactions on the site. If cognitive dissonance explained the findings, we would expect people in the problematic use group to also rate their interactions as more meaningful. An alternative interpretation is that problematic use has both good and bad aspects to it -- a person feels that they get value from Facebook, but may feel overly reliant on it or that they lack control.

The way they spent their time on the site also differed (Figure \ref{fig:cmp_all}b). Compared to the non-problematic use group, people reporting problematic use spent a smaller proportion of their time viewing content on News Feed (-7.7\%, \textit{d} = 0.18, \textit{p} < 0.001), and a greater proportion on profiles, both their own and others' (17.9\%, \textit{d} = 0.15, \textit{p} < 0.01). They were no different in the proportion of their time they spent in Messenger, groups, videos, or Pages (n.s.).

Because of the greater proportion of time people experiencing problematic use spent on profiles, we conducted several post-hoc analyses to better understand if they were using profiles any differently.
People in the problematic use group were no more likely to be viewing their own profile or a friend's (n.s.).
Examining the fraction of profile views coming directly after a previous profile view, they were also no more likely to serially ``hop'' from profile to profile (n.s.).
In addition, as profile pages include a link to message the profile owner, the increased time spent on profiles may be partially due to people messaging others more (we discuss messaging in more detail later in this section).
Indeed, a regression analysis revealed that the number of messages sent, number of friends messaged, and whether a person reported problematic use were all significant predictors of time spent viewing profiles (\textit{p} < 0.01).

\subsection{Connection and Tie Strength}

\subsubsection{Interactions with close friends}
As strong-tie interactions have been associated with improved well-being, we expected that problematic use would be associated with fewer interactions with close friends and more interactions with weak ties.
However, we found that people in the problematic use group were no different than people in the non-problematic use group in terms of the proportion of content they viewed from close friends, their network density, the proportion of close friends' profiles they viewed, and the proportion of messages they sent or received from close friends (Figure \ref{fig:cmp_all}c, Research Question 3). They were also no different in how meaningful they said their interactions on the site were.

Problematic use was also not associated with people seeking out interactions outside of their friend networks: there were no group differences with respect to the frequency of sending friend requests, likelihood of friend requests being accepted, fraction of profile views that were of non-friends, or fraction of comments on non-friend posts (n.s.).

\subsubsection{Synchronous messaging}
While people experiencing problematic use do not spend proportionally more time messaging others, they still sent 62.7\% more messages than those who are not experiencing problematic use (\textit{d} = 0.20, \textit{p} < 0.001) (Figure \ref{fig:cmp_all}d), despite spending only 21.6\% more time overall on Facebook.
Normalizing by the amount of time spent on the site, they sent 38.7\% more messages per hour (\textit{d} = 0.19, \textit{p} < 0.001).
They were also 36.7\% more likely to have sent more messages than they received (\textit{d} = 0.19, \textit{p} < 0.01).
There were no differences in the mean number of words per message they sent or received, suggesting that these differences are not due to longer messages being split up into a series of smaller ones.

Overall, our findings on messaging activity and time spent contradict an image of people experiencing problematic use because of hours of unintentional scrolling or serially watching videos. Instead, they paint a picture of people spending more time browsing profiles and messaging others.

\subsubsection{Feedback received and given}
There were no significant differences between the problematic and non-problematic use groups in the number of likes per post or comments per post that people received, or in the number of likes or comments people gave per post that they viewed (not shown).

\subsection{Control}

\subsubsection{Notifications}
People reporting problematic use received 27.4\% more notifications than people who did not report problematic use (\textit{d} = 0.15, \textit{p} < 0.05), and responded to a greater fraction of these notifications (\textit{d} = 0.18, \textit{p} < 0.01, Figure \ref{fig:cmp_all}e, Research Question 4).
In particular, they were more likely to respond to notifications when they were about replies to comments they had made (\textit{d} = 0.18,  \textit{p} < 0.05).
They did not respond to notifications any more quickly (n.s.).

The correlational data do not allow us to determine if notifications contribute to feelings of problematic use, or if the differences in notification volume and likelihood of responding reflect different levels of engagement and friend activity (e.g., more friends sharing content).

\subsubsection{Deactivation}
People in the problematic use group were 2.6x as likely to have deactivated their accounts in the past four weeks (\textit{d} = 0.16, \textit{p} < 0.05), compared to people in the non-problematic use group (Figure \ref{fig:cmp_all}f, Research Question 5). These deactivations were temporary as respondents had to be using Facebook to be recruited for the survey, so the true number of deactivations among all individuals experiencing problematic use may have been larger. Previous research has described deactivation as a risk-reduction strategy \cite{boyd2010risk} and way to focus \cite{baumer2013limiting}, and that may be the case here: people who feel out of control about their Facebook use may deactivate their accounts to stop notifications, prevent themselves from habitually checking up on friends, or generally take a break from the site.

When people deactivate their accounts, Facebook requires that they provide a reason from a list of options, such as ``I get too many emails, invitations, and requests from Facebook'' or ``I spend too much time using Facebook.'' There were no significant differences between groups in deactivation reasons, suggesting that people have similar reasons for deactivating, even if their use of the site isn't problematic. As we discuss below, designers of social network sites may want to offer more granular controls than deactivation to allow people to better manage their time, prevent interruptions, and break problematic habits.

\subsection{Social Narratives}
Participants experiencing problematic use were 2.1x as likely to have viewed posts and comments about social media or phone addiction (\textit{d} = 0.22, \textit{p} < 0.001, Figure \ref{fig:cmp_all}g, Research Question 6).
On one hand, people who experience problematic use may be more likely to look up content about addiction or know others who also experience problematic use.
On the other hand, people who are exposed to discussion about social media or smartphone addiction may be more likely to think about these problems in their own lives.

\section{Discussion and Conclusion}

\subsubsection{Summary}
Approximately 3\% of Facebook users in the U.S. report feeling like Facebook contributes to problems with their sleep, work, or relationships and that their Facebook use is difficult to control. Understanding their experiences on the platform can help designers develop supportive and context-appropriate tools to reduce negative impact associated with problematic use. This study presents several key differences between people reporting problematic use and those who do not, including greater time spent, particularly late at night; responding to a larger fraction of notifications; spending a greater proportion of time browsing friends' profiles; being more likely to deactivate;  sending more messages than one receives; and reading more content about technology addiction. Demographic factors also play a role: men and younger people were more likely to feel that their use of Facebook was problematic, as were people who had gone through recent major life events such as breakups or moves.

Despite feeling like there were areas of their lives that were negatively impacted by Facebook use, people in the problematic use group also rated Facebook as more valuable in their lives than did people in the non-problematic use group, demonstrating that the technology is not uniformly beneficial or harmful. As designers, we should identify ways to help people avoid problematic use so that people can continue to get that value. People in the problematic group were nearly three times as likely to deactivate their accounts, which suggests they were attempting to gain more control over their time on the site, but deactivation cuts off access to that value. Later, we discuss design implications that may be more useful and flexible than deactivation.

\subsubsection{Major life events}
Major life events such as breakups or moves were associated with higher rates of problematic use, but the causal direction is unclear. Breakups could cause people to use technology in different ways, such as sending more messages to friends seeking support, or surveilling an ex's profile. They could also cause people to view their lives through a lens that makes other activities, such as technology use, seem problematic. Or, a major life event could be associated with a change in routine, and thus could be a vulnerable time for problematic patterns to be strengthened, when people have more time on their hands, are feeling upset or less social, or have something important they want to talk about through social media. Reverse causation is also possible: problematic technology use could lead to major life events; if technology use negatively affects sleep, relationships, or work performance, it could lead to a breakup, job loss, or move. Here, we do find an association between major life events and changes in behavior across both problematic and non-problematic use groups. In our data, major life events predicts more message-sending in both groups (\textit{p} < 0.001). But while major life events do play a role in problematic use, they do not entirely account for the differences in behavior associated with problematic use. People experiencing problematic use still send more messages, even after controlling for whether they had a major life event in the past two months (\textit{p} < 0.001).

\subsubsection{Moderation in use}
As we show in the present study, people who feel like they have a problem are more likely to deactivate their accounts. However, total avoidance may not be the best solution for everyone. People with problematic use are also more likely to report that their use of Facebook is more valuable, and our findings do not support the interpretation this this is due to cognitive dissonance or rationalization. Studies further show that moderate social media use results in more positive well-being outcomes than no social media use \cite{przybylski2017large,twenge2018decreases}.
Nonetheless, while moderate, controlled use may be the most appropriate recommendation for the general population, abstinence from problematic applications may still be warranted for individuals experiencing clinical-level concern with their internet behaviors \cite{petersen2009pathological,young2010internet}.

\subsection{Design Implications}
These findings suggest multiple opportunities for design, not just on Facebook, but communication platforms more generally. First, the data suggest the need to provide people with more granular options than deactivation to take a break from social media. Designers may want to promote alternative options to increase control and provide for uninterrupted time, such as turning off push notifications, especially at bedtime. Because there were no differences in the reasons for deactivation between the problematic and non-problematic use groups, design changes such as these could be relevant and beneficial for all users seeking a temporary break, not only those experiencing problematic use.

\begin{figure}
\includegraphics[width=\columnwidth]{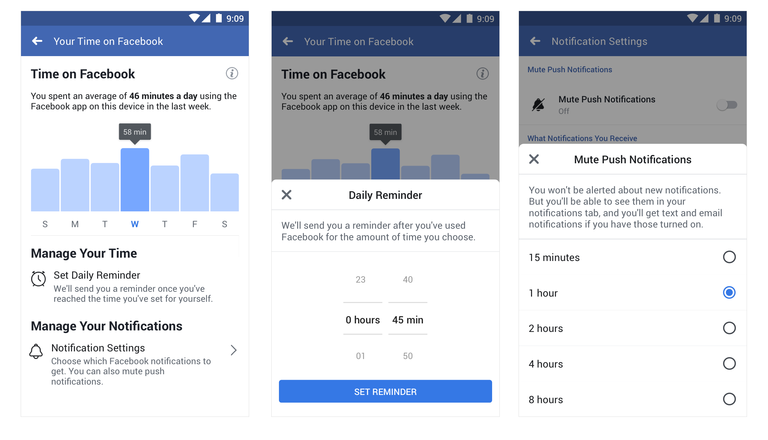}
\caption{Facebook's time management tools provide timers and reminders for people to take a break and options to mute notifications.}
\label{fig:fb_time_management}
\end{figure}

Several technology companies announced new features in 2018 to help people better manage interruptions \cite{apple2018,google2018,facebook2018}. Figure \ref{fig:fb_time_management} shows Facebook's new time management tools, which were informed by this research. The tools include a dashboard to visualize time spent, a time-based reminder to take a break, and options to control or mute notifications.

Additional research is needed to understand what kinds of notifications people find most beneficial, so that designers can better prioritize and filter notifications.  Teens and young adults were more likely to report experiencing problematic use, so designers may want to consider different control settings specifically for younger people.

Problematic use was higher among people experiencing certain major life events, including breakups and moves. These kinds of events are also associated with increases in depression \cite{moreno2011feeling}. Social media platforms could provide additional support for managing these life transitions.

\subsection{Limitations and Future Research}

This method of pairing cross-sectional survey and behavioral data has several limitations. The analysis is correlational rather than causal; we can only report associations between perceived problematic use and site activities but do not know whether those activities cause feelings of problematic use, whether a person's propensity for problematic use causes those activity patterns, or whether something else like a major life event causes both perceptions of problematic use and site activity. We make design recommendations that we hypothesize will have positive outcomes, but further research is necessary to understand their impact on problematic use.

Though the present data come from Facebook, other smartphone apps and communication platforms may present similar opportunities to study problematic use. Notifications, browsing feeds of content, and channels for messaging are common across platforms. However, platforms differ in network composition, communication synchronicity, media type (e.g., images versus text posts), and motivations for use. For instance, some studies suggest that visual media provide more gratification than text media, and so may be perceived to be more compelling for increased use \cite{dunne2010young}. How these differences relate to problematic use remains future work.

More research is also necessary to understand how other factors may contribute to problematic use. For example, the popularity of social media and instant messaging has created pressure to always be available, particularly among youth \cite{greenfield2011addictive}. While we found mixed evidence for this---people reporting problematic use responded to a larger fraction of notifications but did not respond any quicker---additional qualitative work could probe more deeply into the connection between availability expectations and problematic use. Upward social comparison may also lead to problematic use, especially among teens as peer influence is much stronger in adolescence than in adulthood \cite{steinberg2007age}.

Additional work is also necessary to better interpret the differences that we observed.
For instance, though we found that people reporting problematic use spent proportionally more time viewing profiles, profile viewing is associated with both positive and negative outcomes.
People who use profile pages more may be using Facebook primarily to keep up with friends they do not see as often, and this greater awareness of what others are doing can increase feelings of closeness \cite{burke2014growing}.
However, spending time viewing profiles of acquaintances also lowers self-esteem \cite{vogel2015compares} (though viewing one's own profile instead increases self-esteem \cite{gonzales2011mirror}).
Understanding the causal pathway between profile viewing and problematic use, if any exists, remains future work.

Several methodological limitations exist.
For example, we only log time spent when the Facebook app is active on a person's screen, but do not know if they are looking at the screen the whole time; our measures of closeness may not necessarily identify every individual's closest friends.
There are also selection biases among survey participants. People who have permanently quit Facebook are missing from the sample, so our statistics related to account deactivation are likely underestimates. We also do not know the relationship between problematic use and account deletion. Surveys outside of Facebook (e.g., \cite{baumer2013limiting}) are useful for understanding the motivations of people who have left the platform permanently. There may be other sources of response bias: people who stay up late may be more willing to complete surveys. The data are U.S.-centric and reports of problematic use and associated behaviors may differ internationally based on cultural differences, mobile broadband adoption, and norms. For example, time spent on social media varies by country. In 2017, people in the Philippines spent almost twice as much time on social media as people in the U.S. \cite{we2018social}. Additional international research is needed.

This research was quantitative. While it includes granular information about the kinds of activities that are associated with problematic use, we need additional qualitative research to better understand why people with problematic use engage with technology in the ways that they do and what would best help them gain control.

These challenges related to problematic use are not specific to Facebook---many of the findings in the present study generalize to other social media and smartphone technology more broadly. As researchers and designers we should continue to address the serious challenges that people face with technology in order to ensure it best serves its role in supporting people's lives.

\begin{acks}
The authors would like to thank Bethany de Gant, Jennifer Guadagno, Alex Dow, Lada Adamic, and Robert Kraut for feedback and assistance with this work.
\end{acks}

\balance{}
\bibliographystyle{ACM-Reference-Format}
\bibliography{references}

\end{document}